\pdfoutput=1
\documentclass[11pt,twoside]{article}
\usepackage{graphicx,epsfig,natbib,epstopdf}
\usepackage{CS18}
\usepackage{amsfonts,amsmath,amssymb}
\usepackage{url}
\defcitealias{1972ApJ...171..565S}{S72}

\begin{document}

\title{Upgrading the Solar-Stellar Connection: News about activity in Cool
Stars}

\author{H. M. G\"unther$^{1}$, K. Poppenhaeger$^{1,10}$, P. Testa $^{1}$,
S. Borgniet$^2$, 
S. Brun$^3$,       
H. Cegla$^4$,      
C. Garraffo$^1$,
A. Kowalski$^5$,   
A. Shapiro$^6$,    
E. Shkolnik$^7$,   
F. Spada$^8$,      
A. Vidotto$^9$}    
\affil{$^1$Harvard-Smithsonian Center for Astrophysics, 60 Garden Street,
  Cambridge, MA 02138, USA}
\affil{$^2$Institut d'Astrophysique et de Plan\'etologie de Grenoble, CNRS-UJF UMR 5571, 414 rue de la Piscine, 38400 St Martin d'H\`res, France}
\affil{$^3$Laboratoire AIM Paris-Saclay, CEA/Irfu Universit\'e Paris-Diderot CNRS/INSU, F-91191 Gif-sur-Yvette, France}
\affil{$^4$Astrophysics Research Centre, School of Mathematics \& Physics, Queen's University, University Road, Belfast BT7 1NN, UK}
\affil{$^5$NASA-GSFC, Heliophysics Science Division, 8800 Greenbelt Rd, Greenbelt, MD 20771,USA}
\affil{$^6$Physikalisch-Meteorologisches Observatorium Davos, World Radiation centre, 7260 Davos Dorf, Switzerland}
\affil{$^7$Lowell Observatory, 1400 West Mars Hill Road, Flagstaff, Arizona, USA}
\affil{$^8$Leibniz-Institut f\"ur Astrophysik Potsdam (AIP), An der Sternwarte 16, D-14482, Potsdam, Germany}
\affil{$^9$Observatoire de Gen\`eve, Universit\'e de Gen\`eve, Chemin des Maillettes 51, Versoix, 1290, Switzerland}

\affil{$^{10}$NASA Sagan fellow}

\begin{abstract}
In this splinter session, ten speakers presented results on solar and
stellar activity and how the two fields are connected. This was followed
by a lively discussion and supplemented by short, one-minute highlight
talks. The talks presented new theoretical and observational results on
mass accretion on the Sun, the activity rate of flare stars, the
evolution of the stellar magnetic field on time scales of a single cycle
and over the lifetime of a star, and two different approaches to model
the radial-velocity jitter in cool stars that is due to the granulation
on the surface. Talks and discussion showed how much the interpretation
of stellar activity data relies on the sun and how the large number of
objects available in stellar studies can extend the parameter range of
activity models.

\end{abstract}

\section{Introduction}

Coronal magnetic activity is what makes Cool Stars special and distinguishes them from hotter stars without magnetic dynamos. Coronal activity also shapes the environment of many extrasolar planets, giving rise to a completely new perspective why we care about magnetic phenomena. While coronal magnetic activity has been studied for a long time, there are new and exciting insights from recent solar and stellar missions, and this session provided a forum to bring the stellar and solar astronomers together and discuss synergies from both fields.

In two key areas there is significant observational progress in the last few years:

\begin{itemize}
\item New solar observations, especially IRIS (Interface Region Imaging Spectrograph) and SDO/AIA (Solar Dynamics Observatory, Atmospheric Imaging Assembly) have significantly improved the spatial and temporal resolution of solar observations, as well as their spatial/temporal/temperature coverage.

\item The sample size of stellar activity surveys has exploded, specifically due to the data from Kepler and COROT. We are now at a crucial stage to understand and recap the wealth of such observations, especially in the light of future high-energy missions like Athena+, which was recently selected by ESA for the L2 launch opportunity.
\end{itemize}

There is an intimate connection between the activity of our Sun and cool stars. Almost all interpretation of coronal activity starts from the Sun, where observations with a spatial and temporal resolution that is unmatched in stellar astrophysics are possible. On the other hand, active stars provide for more energetic events than commonly observed in our Sun, so that the stellar perspective can help to extend models to a wider range, e.g. to explain the so-called super-flares. 

This splinter session consisted of ten major talks that presented both observational and theoretical studies of stellar activity. The content of these presentations will be summarized in the following sections. Each presentation was followed by a lively debate with the audience. In addition, several short, one-minute presentations allowed conference attendees to highlight relevant posters or advertise other work.
Slides from the pretensions are can be retrieved from \href{https://zenodo.org/collection/user-coolstar18splitersolarstellarconnection}{Zenodo} or from \href{http://www.coolstars18.net/}{the conference website} if the speakers decided to make them available.

In addition to this summary, presentations given in this splinter session may also be discussed as separate articles in this volume.

In the following sections, we summarize the results presented in this splinter session, roughly ordered by the timescale of the variability phenomenon, although all these phenomena are of course related. Section~\ref{sect:sun} presents infall on the Sun as a template for accretion in young stars, Section~\ref{sect:flares} looks at stars with a high occurrence rate of flares, Section~\ref{sect:cycles} presents studies of phenomena that change over the solar or stellar cycle, such as the stellar radius, angular momentum loss, and activity level, and Section~\ref{sect:evolution} describes changes of the activity level over the lifetime of a star. Finally, we summarize this splinter session in Section~\ref{sect:summary}

\section{The sun as a template for accreting stars}
\label{sect:sun}
Paola Testa opened the splinter session with a view towards the sun as a template for accreting, young, low-mass stars, the so called classical T~Tauri stars \citep[CTTS -- see review by][in the proceedings of Cool Stars 17]{2013AN....334...67G}. CTTS are surrounded by an accretion disk and material falls from this disk onto the star. It is accelerated to almost free-fall velocities of a few hundred km~s$^{-1}$ and causes a strong shock wave when it impacts the stellar surface. According to our current understanding of CTTS this accretion shock produces X-ray and UV emission lines and heats the surrounding stellar atmosphere causing an additional continuum emission, the so-called veiling. In CTTS, we cannot resolve the stellar surface, so the details of this process are not well understood. 

Paola Testa and her collaborators realized that the sun can provide a template for these processes, when erupted fragments fall back onto the sun and cause bright, hot impacts. They observed several impacts of falling fragments after the eruption of a filament in a solar flare on 7 June 2011. As imaged in the ultraviolet (UV)-extreme UV range by the Atmospheric Imaging Assembly \citep{2012SoPh..275...17L} on board the Solar Dynamics Observatory, many impacts of dark, dense matter display uncommonly intense, compact brightenings. High-resolution hydrodynamic simulations show that such bright spots, with plasma temperatures increasing from $10^4$ to $10^6$~K, occur when high-density plasma ($>10^{10}$ particles per cubic centimeter) hits the solar surface at several hundred kilometers per second, producing high-energy emission as in stellar accretion. The high-energy emission comes from the original fragment material and is heavily absorbed by optically thick plasma, possibly explaining the lower mass accretion rates inferred from X-rays relative to UV-optical-near infrared observations of young stars. This work is published in \citet{2013Sci...341..251R} and provides a good example how we can use the high spatial resolution of solar observations to enhance our understanding of other stars.

\section{Flares}
\label{sect:flares}

Adam Kowalski presented work on stellar flares. Flares are the most violent events in a stellar atmosphere and they are also fast. The typical duration of a flare on a dMe star is only a few minutes \citep{2003ApJ...597..535H}, although long flares can last for several hours. For the most active stars, the flare rate can be several orders of magnitude higher than on the sun. Despite these differences, we usually interpret flares from all stars analogous to solar activity. A.~Kowalski presented a deep rapid archival flare transient search in the galactic bulge, published by \citet{2012ApJ...754....4O}. 
This talk focused on the properties of flares and flare stars in the optical as revealed by a Hubble Space Telescope/ACS planet search of the Galactic Bulge, which monitored over 200,000 dwarf stars over seven days at a cadence of just a few minutes. This project discovered $\approx100$ flare stars. The occurrence rate for flares on those stars is about three orders of magnitude above the solar rate; also, most flare stars have rotation periods below three days. This could be an indication that the flaring stars represent a population of old (10 Gyr), close binaries that have spun-up over their lifetime because tidal forces tend to synchronize the stellar rotation with the orbital period.

Given this high number of flare stars, we can expect much larger samples of flares in future transient searches, e.g.\ LSST will have 50-100 such flares in any single exposure in the galactic plane.

\section{Solar and stellar cycles}
\label{sect:cycles}
Federico Spada presented work on radius changes of solar-like stars over their activity cycle. Low mass stars ($M\lesssim 1 \, M_\odot$) have a subsurface convection zone of sufficient thickness (i.e., a few percent of the total mass) to host a dynamo, which sustains the generation of large-scale magnetic fields. 
Although the details of this process are far from established, even for the Sun, a common feature of all dynamo models is that the magnetic field will be generated and transported through the convection zone, changing in configuration, strength, and depth during the magnetic cycle.
Magnetic fields affect the stellar structure in a variety of ways, including both direct (e.g., by contributing an extra pressure term to the hydrostatic equilibrium) and indirect effects (e.g., by inhibiting or suppressing convective motions). 
As a consequence, it can be expected that the interior structure and the global parameters of the stars (i.e., radius, luminosity, effective temperature) will change during the cycle.

From space-based measurements, the Total Solar Irradiance (TSI) of the sun is known to vary by about $0.1 \%$ within a cycle \citep{2013SSRv..176..237F}. 
Since the TSI can vary because of surface phenomena (e.g. sunspots, faculae, ...) as well as intrinsic luminosity changes, the TSI measurement provides an upper limit to the variation of $L$. 
Moreover, a variation of the solar radius of about $2\cdot 10^{-4}$ within a cycle has been recently reported by \citet{2013MNRAS.436.2151S}.
Finally, from the analysis of the helioseismic frequency splittings, it is possible to place an upper limit on the strength of the field of $0.4$--$20$ kG and of $0.3$--$1$ MG in outer the layers and near the bottom of the convection zone, respectively \citep{1993ASPC...42..229G,2002ApJ...578L.157C,2000JApA...21..343A,2009ApJ...705.1704B}.

Spada and coworkers have modified the Yale Rotational stellar Evolution Code (YREC), including up-to-date input physics \citep{2008Ap&SS.316...31D}, to take into account the effect of magnetic fields, according to the formulation of \citet{1995ApJS..101..357L}.
The approach followed is perturbative in nature, and is therefore applicable to two extreme cases: to young, fast rotators, hosting strong magnetic fields, but where the precision required by the comparison with available observations is not extremely high \citep[see, e.g., the work by][]{2012ApJ...757...42F,2013ApJ...779..183F,2014ApJ...789...53F}; and to the Sun, whose magnetic-induced variability is very small.  

They have applied their version of YREC, implementing the treatment of magnetic effects on the stellar structure, to the solar case. 
Their results (Spada et al., in preparation; for earlier results, see \citep{Spada2013}) show that it is possible to reproduce the observed radius variation of $\delta R \approx 2\cdot 10^{-4}$, satisfying the upper limits on the luminosity ($\delta L \le 10^{-3}$) and on the magnetic field strength. 

\begin{figure}[h!]
\begin{center}
\includegraphics[width=0.5\textwidth]{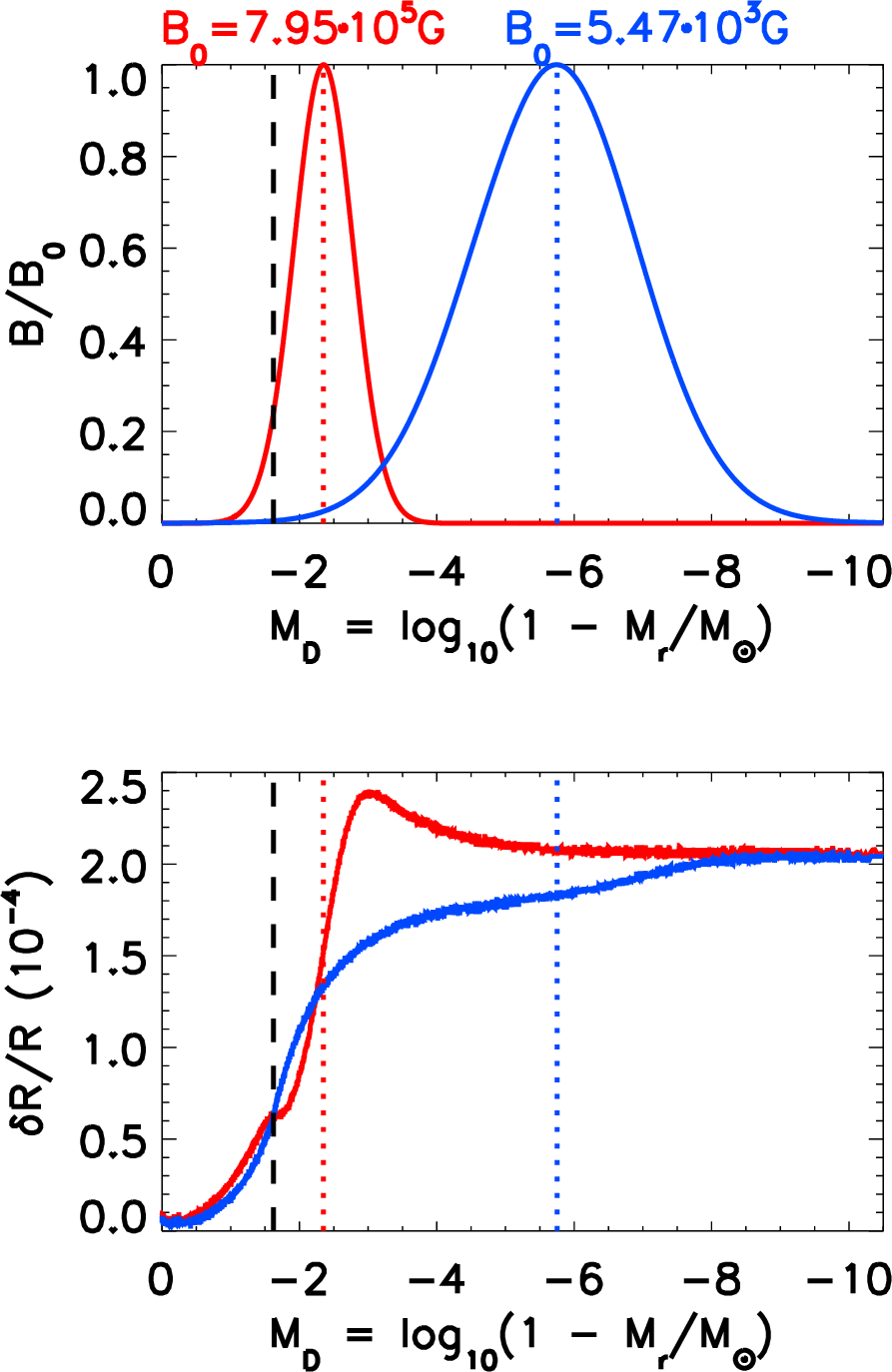}
\caption{\label{fig:solarinterior}
Radius variation in the interior of a solar model induced by a magnetic perturbation peaking near the bottom (red) and the outer layers (blue) of the convection zone. \emph{Upper panel:} radial profiles of the magnetic field, as a function of logarithmic mass depth, $M_D$. \emph{Lower panel:} shell-by-shell radius variations, relative to a model without magnetic fields. The vertical dashed line marks the base of the convection zone.}
\end{center}
\end{figure}

The radius variation in the interior with respect to a model without magnetic perturbation is shown in Figure~\ref{fig:solarinterior}. 
The Figure shows two illustrative cases, characterized by a Gaussian-like radial profile of the magnetic field, peaking near the bottom of the convection zone ($r=0.71$) and in the outer layers (around $r=0.9$).
In both cases, the peak field strength, $B_0$, has been adjusted to obtain $(R_{\rm pert} - R_0)/R_0 \approx 2\cdot 10^{-4}$.

Cecilia Garraffo then presented work on angular momentum loss. Stars spin-down through their magnetized winds that carry away mass and angular momentum (magnetic braking). These winds
corotate with them up to a certain distance, called Alfv\'en radius ($R_A$), where the speed of the wind reaches the Alfv\'enic speed ($ u_A=B/\sqrt{4\pi \rho}$, with $B$ the magnetic field strength and $\rho$ the density of the plasma).  It is widely believed that stellar spin-down is a function of magnetic activity only.  This has its basis on the famous expression by \citet{Weber67} $\dot{J}=\frac{2}{3} \Omega \dot{M}R_A^2 $, where $J$ is the angular momentum, $\Omega$ the angular velocity, $M$ the mass, and where spherical symmetry has been
assumed ($R_A$ is constant), making it a good first order approximation.  However, for a more
realistic approach one should take the magnetic topology on the surface of the star into account, since it plays an important role in determining the efficiency of magnetic braking \citep{Cohen09,Garraffo13,Lang14}.  Using a three dimensional MHD, self-consistent, physics-based code, {\it BATS-R-US} \citep{Powell99,Toth12}, Garraffo and coworkers performed simulations to explore the effect of magnetic active regions on mass and angular momentum loss rates. 

\begin{figure}[h!]
\begin{center}
\plotone{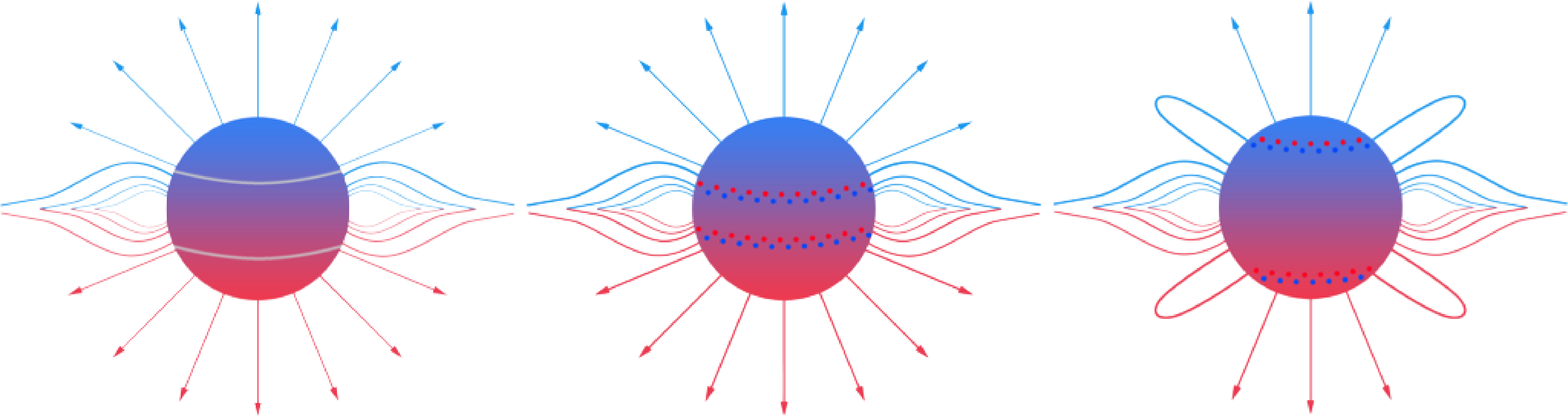}
\caption{\label{fig:1}
Qualitative plot of winds structure for a dipole (left), and the same dipole with low-latitude (center) and high-latitude (right) magnetic active regions. The limiting latitude between the open and closed lines regime of the dipole is plotted in white (left).}
\end{center}
\end{figure}

\begin{figure}[h!]
\begin{center}
\plotone{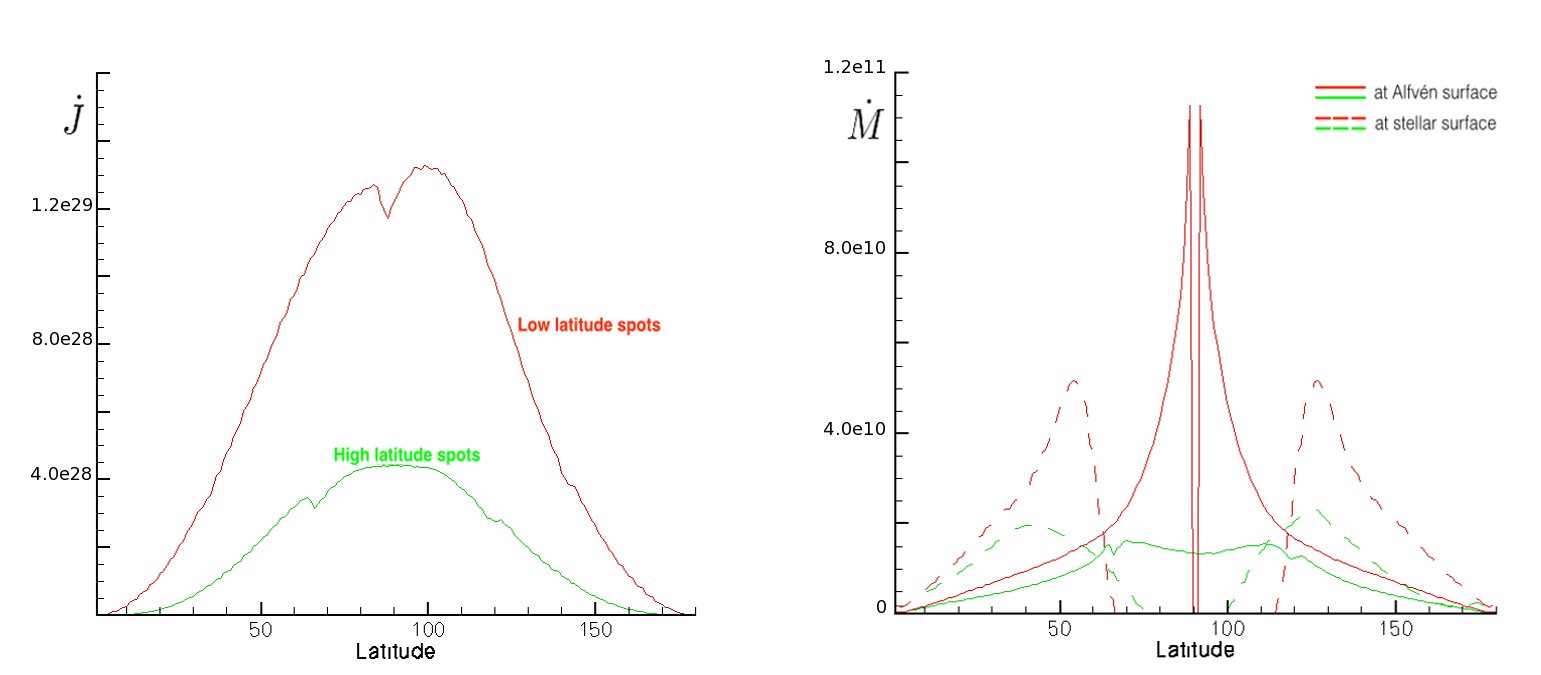}
\caption{\label{fig:2}
Mass and angular momentum loss as a function of latitude at the Alfv\'en Surface (solid line) and at the stellar surface (dashed line) for low-latitude (red) and high-latitude magnetic active regions.}
\end{center}
\end{figure}

They find a bi-modal regime regulated by the latitude of spots by which magnetic active regions efficiently reduce mass and angular momentum loss rates {\it only} when located outside of the dead zone (closed field lines region on the stellar surface).  The mechanism behind it is the closing of otherwise open lines that leads to a reduction in the plasma being carried away (see Fig~\ref{fig:1}).  As a result, magnetic cycles of stars whose spots cross the limiting latitude of the dead-zone, and therefore turn on and off this mechanism,  will experience a large modulation of mass and angular momentum loss, typically of the order of a few (see Fig~\ref{fig:2}).  

A. Shapiro presented recent results on the modeling of photometric variability of sunlike stars. The Sun and stars with low magnetic activity levels become photometrically brighter when their activity increases. Magnetically more active stars display the opposite behavior and get fainter when their activity increases. Shapiro and coworkers reproduced the observed photometric trends in stellar variations with a model that attributes the variability of the stellar radiative energy flux to the imbalance between the contributions from dark starspots and bright faculae \citep{Shapiro2014arXiv}. Their approach allowed them to model the stellar photometric variability vs. activity dependence and reproduce the transition from faculae-dominated variability and direct activity-brightness correlation to spot-dominated variability and inverse activity-brightness correlation with increasing chromospheric activity level. The ability of the model to reproduce the behavior of Sunlike stars is an indication that the photometric variability of more active stars has the same basic causes as the Sun's.

Simon Borgniet showed recent work on a model for solar-like activity patterns. Future high-resolution spectrographs are expected to push back the radial velocity detection limits at the level of a few cm/s, giving theoretically access to low planetary masses such as Earth-like planets. However, according to many studies, stellar magnetic activity will induce radial velocity jitter at the level of the m/s, i.e.\ far over the expected performances of the future spectrographs, even in the case of a low activity star. It will thus seriously undermine the possibility of detecting an Earth twin in the habitable zone of its host star, unless precise correction can be done. In the context of the modeling of such stellar jitter, Borgniet and coworkers presented a fully parametrized model of the activity pattern of a Solar-like star and of its impact on radial velocity jitter. The model includes dark spots, bright faculae and the attenuation of the convective blueshift. It has been compared to the Solar pattern over a full Solar cycle for validation. Being fully parametrized, it is straightforward to transfer to other spectral types and stellar properties, which is a work in progress. It will allow to predict the radial velocity signature for a wide range of stars and activity levels. It also opens new perspectives in terms of correcting the activity radial velocity signature.

Heather Cegla presented a model for astrophysical noise in planet detections based on magneto-convection on the stellar surface. Cool, low mass stars with a convective envelope have bubbles of hot, bright plasma rising to the surface where they eventually cool, darken and sink. The motions of these plasma bubbles induce stellar line asymmetries since the radial velocity (RV) shift induced from the uprising granules does not completely cancel the shift from the sinking intergranular lanes. Furthermore, these line asymmetries are constantly changing as the ratio of granular to intergranular lane material continues to change due to magnetic field interplay. The net result for Sun-like stars is shifts in the line profiles on the order of several tens of cm/s. Hence an understanding of magneto-convection and its effects is paramount in any high precision RV study. As mentioned in the last paragraph, one particular area impacted is the RV confirmation of Earth-analogs; the astrophysical noise from the host star stellar surface magneto-convection completely swamps the 10 cm/s signal induced from the planet. Cegla and coworkers aimed to understand the physical processes involved in order to disentangle the effects of magneto-convection from observed stellar lines. To do so, they started with a state-of-the-art 3D magnetohydrodynamic simulation of the solar surface \citep{Cegla2013}. Motivated by computational constraints and a desire to breakdown the physics, they parameterized the granulation signal from these simulations. This parameterization was then used to construct model Sun-as-a-star observations with a RV precision far beyond current instrumentation. This parameterization across the stellar disc, for a variety of magnetic field strengths, was presented here, alongside the current results from the model star observations. They found several line characteristics to be correlated with the induced RV shifts. Particularly high correlations were found for the velocity asymmetry (comparing the spectral information content of the blue wing to the red wing) and brightness measurements (approximated by integrating under the model observation profiles), allowing significant granulation noise reduction.

This talk was really complementary to the work of Borgniet and co-workers as it tackled the same problem with a different approach. At this point it is not clear, if an empirical correction or a simulation-based model will provide a better correction for the RV jitter do to stellar granulation and activity. The results of both campaigns can feed directly into future high precision RV studies, such as the search for habitable, rocky worlds, with the forthcoming ESPRESSO and E-ELT/HIRES spectrographs.

Sacha Brun presented a model for a spot-dynamo for solar-like stars. He started with an overview about how large scale flows are established, how they vary with rotation rate and how this impacts dynamo action in stars. He then discussed the concept of a spot-dynamo, which is a nonlinear dynamo generating self consistently rising omega-loops. In particular, their simulations show that magnetic wreath-like structures can become turbulent and intermittent enough so that intense bundles of fields reach 50 kG and start becoming buoyant, forming omega-loop like structures \citep{Nelson2013}. Their simulations aim to improve current models in order to achieve more realistic models for stellar magnetic activity.

\section{Activity evolution over the livetime of a star}
\label{sect:evolution}
The activity of cool stars evolves from their pre-main sequence origins all the way through the main sequence; in fact, the change in magnetic activity is one of the most pronounced effects of aging and thus it can be used as a rough tool to estimate the age of a star (``gyrochronology''). 
In this area of astrophysics the information that we receive from the sun is limited, since we observe it only at one point in its evolution and only limited data is available about long-term changes in the solar activity. Instead, we need to turn to samples of stars with different ages to understand how activity changes and we can then use this knowledge to predict the future evolution of our own Sun.

As discussed by Cecilia Garraffo, magnetic fields play an important role in regulating stellar rotation, from the early stages of star formation to the death of stars. At the main-sequence (MS) phase, `isolated' stars (single stars and stars in multiple systems with negligible tidal interaction) slowly spin down as they age. This fact was first observed by \citet[][S72, from now on]{1972ApJ...171..565S}, who empirically determined that the projected rotational velocities $v \sin(i)$ of G-type stars in the MS phase decrease with age $t$ as $v\sin(i) \propto t^{-1/2}$. This relation, called the ``Skumanich law'', has served as the basis of the gyrochoronology method \citep{2003ApJ...586..464B}, which yields age estimates based on rotation measurements. The rotational braking observed by \citetalias{1972ApJ...171..565S} is believed to be caused by stellar winds, which, outflowing along magnetic field lines, are able to efficiently remove the angular momentum of the star \citep[e.g.,][]{1958ApJ...128..664P,1962AnAp...25...18S} and indicators of magnetic activity, such as surface spot coverage, emission from the chromosphere, transition region or corona, have been recognized to be closely linked to rotation. 

Two talks in this splinter session presented new results about this long-term evolution of stellar activity. \citet{2014MNRAS.441.2361V} investigated how the large-scale surface magnetic fields of cool dwarf stars, reconstructed using the Zeeman Doppler imaging (ZDI) technique\footnote{This magnetic field-measuring technique consists of analyzing a series of circularly polarized spectra (Stokes V signatures) to recover information about the large-scale magnetic field (i.e., its intensity and orientation; \citealt{1997A&A...326.1135D}). Since the ZDI technique measures the magnetic flux averaged over surface elements, regions of opposite magnetic polarity within the element resolution cancel each other out \citep{2010MNRAS.404..101J,2011MNRAS.410.2472A}. As a consequence, the ZDI magnetic maps are limited to measuring large-scale magnetic field. }, vary with age $t$, rotation period $P_{\rm rot}$, Rossby number $\rm Ro$ and X-ray luminosity (an activity index) and E.~Shkolnik traced the UV flux of M dwarfs based on \emph{GALEX} observations with a particular view on the consequences this UV bombardment could have on the conditions on planets in the habitable zone.

\begin{figure}[h!]
\begin{center}
\plotone{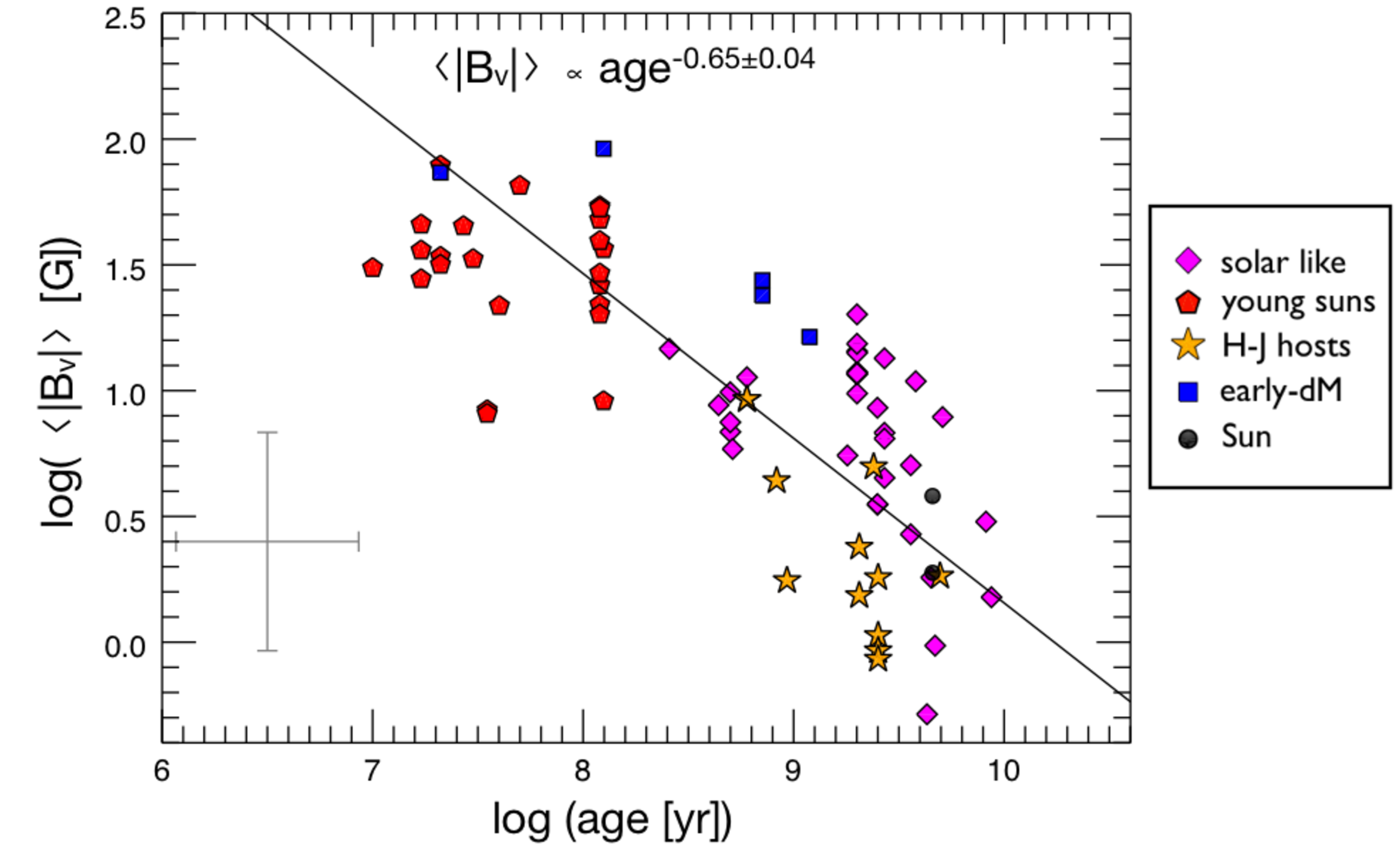}
\caption{\label{fig:vidotto}
The unsigned average large-scale surface magnetic field $\langle |B_V| \rangle$ is related to age as {$t^{-0.65 \pm 0.04}$} (solid line) and  has a similar power-dependence as the Skumanich law. This magnetism-age relation could be used as a way to estimate stellar ages ({``magnetochronology''}), although it would not provide better precision than most of the currently adopted age-dating methods. }
\end{center}
\end{figure}

The sample of \citet{2014MNRAS.441.2361V} consisted of $104$ magnetic maps of $73$ stars, from accreting pre-MS to MS objects, spanning ages from $\sim 1$ Myr to $\sim 10$~Gyr. For the non-accreting dwarfs, they empirically found that the unsigned average large-scale surface magnetic field $\langle |B_V| \rangle$ is related to age as {$t^{-0.655 \pm 0.045}$}, which has a similar power-dependence as the Skumanich law (Figure~\ref{fig:vidotto}). This magnetism-age relation could be used as a way to estimate stellar ages ({``magnetochronology''}), although it would not provide better precision than most of the currently adopted age-dating methods. 

Theoretically, \citetalias{1972ApJ...171..565S}'s relation can be explained on the basis of the simplified wind model of \citet{1967ApJ...148..217W}, further assuming that a linear dynamo (i.e., $B \propto P_{\rm rot}^{-1}$) is in operation. In order to investigate the presence of a linear-type dynamo,  \citet{2014MNRAS.441.2361V} searched for correlations between $\langle |B_V| \rangle$ and rotation. They found that  {$\langle |B_V| \rangle \propto P_{\rm rot}^{-1.32 \pm 0.14}$ or, similarly, $\langle |B_V| \rangle \propto \rm Ro^{-1.38\pm 0.14}$}, which supports the presence of a linear-type dynamo in operation in dwarf stars at the MS phase. 

Additionally, \citet{2014MNRAS.441.2361V} compared the trends found for large-scale stellar magnetism from ZDI studies with similar trends using Zeeman broadening measurements, a complementary technique to measure stellar magnetism that is sensitive to the unsigned large- and small-scale magnetic field $\langle |B_I| \rangle$. They found that trends between $\langle |B_V| \rangle$, or $\langle |B_I| \rangle$, and rotation are roughly similar, indicating that the fields recovered from both techniques might be coupled to each other and that small- and large-scale fields could share the same dynamo field generation processes. 

Another way to empirically constrain the dynamo action in cool stars are
observations in the X-ray or UV regime, where the photospheric contribution to
the total emission is low and the transition region, the chromosphere and the
corona --atmospheric layers that are powered by magnetic activity-- dominate
the total emission. This is true particularly in M dwarfs which have the lowest
photospheric temperatures of all cool stars. M dwarfs are also favorite targets
for habitability studies, because their low luminosity restricts the habitable
zone to small radii, which in turn means that habitable planets can be observed
more easily then around more massive stars. On the other hand, the relatively
more intense FUV radiation influences the chemistry and stability of the
planetary atmosphere
\citep{1993Icar..101..108K,2010Icar..210....1L,2010AsBio..10..751S}.

In this context, \citet{2014arXiv1407.1344S} begun the HAZMAT (HAbitable Zones and M dwarf Activity across Time) program by first measuring the drop in near-UV and far-UV flux in early M stars from 10~Myr to several Gyr using photometry from NASA's Galaxy Evolution Explorer (GALEX). They focus this study on the confirmed low-mass members of nearby young moving groups, the Hyades cluster, and old field stars and show a relatively slow decline in UV flux up until at least 650 Myr with a sharper drop in the old M dwarfs. Yet without confirmed M dwarfs in nearby star clusters with ages of 1-2 Gyr, mapping the precise evolution at these older ages is not currently possible. The UV data also provide much-needed constraints to M dwarf upper-atmosphere models, which are currently insufficient for predicting UV emission from M dwarfs. This analysis will aid empirically motivated upper-atmospheric modeling for the young and old M stars, which can then be used to predict the extreme-UV fluxes most critical to the evolution of a planetary atmosphere. The HAZMAT program is the first comprehensive study of the UV history of M stars.

\section{Summary}
\label{sect:summary}
Large surveys in different wavelength bands have dramatically increased the sample size to study stellar activity such that hundreds of active objects can be identified out of a few hundred thousand target stars. Two examples in the presentations are HST monitoring of the galactic plane and the GALEX all-sky survey but others missions that also contributed a lot of new data to stellar activity (Kepler, CoRoT) were discussed in a parallel splinter session. In addition to lightcurves, A.~Vidotto's ZDI survey probes the magnetic field strength of active stars directly. At the same time, simulations have advanced enough to treat the magnetic field in three dimensions, so that the total stellar angular momentum loss can be calculated for realistic field geometries and disturbances in the interior of the Sun can be linked to a global radius change at the surface.

This splinter session has shown once more how closely our understanding of cool star activity is linked to our understanding of the Sun. For all theoretical models of solar and stellar activity and solar and stellar dynamo theory the sun is the most important benchmark, since it provides the largest set of observational constraints. It is critical that simulations for cool stars can reproduce the solar behavior to gain confidence that they describe the right physics before they can be used to interpret data from other stars. Observationally, the data we receive from the sun is so detailed, that we can even use it as a model system to understand processes such as mass accretion from a disk, although our sun has no accretion disk at all. We thus have to make an effort as a community that solar physicist and astrophysicists engage into a fruitful dialog. The workshops on ``Cool stars, stellar system and the Sun'' provide one venue to do so and the large interest in this splinter session shows that there is a need to discuss these topics. Future workshops of this series should strengthen the effort to include the heliophysics community to increase the participatation in this field.

\end{document}